\documentclass[12pt]{article}

\usepackage{amssymb}
\usepackage{amsmath}
\usepackage{amsfonts}

\newcommand{\R}{\mathbb{R}}
\newcommand{\C}{\mathbb{C}}

\newcommand{\be}{\begin{equation}}
\newcommand{\bea}{\begin{eqnarray}}
\newcommand{\eea}{\end{eqnarray}}
\newcommand{\nn}{\nonumber}
\newcommand{\kt}{\rangle}
\newcommand{\br}{\langle}

\newcommand{\ed}{\end{document}}
\newcommand{\bbr}{\br\!\br}
\newcommand{\kkt}{\kt\!\kt}
\newcommand{\cbr}{(\!(}
\newcommand{\ckt}{)\!)}

\begin{document}

\title{Probability Interpretation for Klein-Gordon Fields and the Hilbert Space Problem in Quantum Cosmology}
\author{Ali Mostafazadeh\thanks{E-mail address: amostafazadeh@ku.edu.tr}\\ \\
Department of Mathematics, Ko\c{c} University,\\
Rumelifeneri Yolu, 34450 Sariyer, Istanbul, Turkey}
\date{ }
\maketitle

\begin{abstract}
We give an explicit construction of a positive-definite invariant
inner-product for the Klein-Gordon fields, thus solving the old
problem of the probability interpretation of Klein-Gordon fields
without having to restrict to the subspaces of the
positive-frequency solutions. Our method has a much wider domain
of application and may be used to obtain a general class of
invariant inner-product on the solution space of a broad class of
Klein-Gordon type evolution equations. We explore its consequences
for the solutions of the Wheeler-DeWitt equation associated with
the FRW-massive-real-scalar-field models.
\end{abstract}


\section{Introduction}

The birth of modern particle physics and the advent of
relativistic quantum field theories have their origin in Dirac's
attempts to obtain a consistent probability interpretation for
Klein-Gordon fields. These attempts led to the discovery of the
Dirac equation, Dirac's theory of holes, and the positron on the
one hand, and the formulation of the method of second quantization
on the other hand. Yet, these developments did not provide a
satisfactory resolution of the issue of the probability
interpretation for Klein-Gordon fields. Because first quantized
scalar fields did not play an important role in high energy
physics this problem did not attract much attention until the
1960s when John Wheeler and Bryce DeWitt founded quantum
cosmology.

One of the outstanding problems of quantum cosmology is the issue
of how to interpret the wave function of the universe
\cite{bryce-1,page,kuchar,isham,wiltshire,carlip}. This is a
scalar field satisfying the Wheeler-DeWitt equation. The latter
has the structure of a Klein-Gordon equation and is plagued among
other things with the problem of negative probabilities, if one
adopts the invariant but indefinite Klein-Gordon inner product
\cite{bryce-1,vilenkin,wald}. Use of the non-invariant $L^2$-inner
product allows for a conditional probability interpretation
\cite{page-w,page}. But this approach does not lead to a
satisfactory resolution of the problem either \cite{kuchar,isham}.
Today, the only successful attempt to attack this problem consists
of Woodard's \cite{woodard} proposal of gauge-fixing the
Wheeler-DeWitt symmetry and its variations known as the method of
refined algebraic quantization and group averaging \cite{marolf}.
These could however be applied to certain special models
\cite{marolf}.

The basic problem of the construction of invariant
positive-definite inner product on the solution space of the
Klein-Gordon type evolution equations has been open for the past
75 years. The purpose of this paper is to report on an application
of the theory of pseudo-Hermitian operators
\cite{p1,p2,p3,p4,jmp-03} that leads to a complete solution of
this problem. In the following we first recall the basic
properties of pseudo-Hermitian Hamiltonians and then discuss their
relevance to the problem of the probability interpretation of the
Klein-Gordon and Wheeler-DeWitt fields. A more detailed treatment
that applies to more general Klein-Gordon type equations is
presented in \cite{cqg}.

\section{Basic Mathematical Results}

A linear operator $H:{\cal H}\to{\cal H}$ acting in a Hilbert
space ${\cal H}$ is said to be pseudo-Hermitian \cite{p1} if there
is a linear, Hermitian, invertible operator $\eta:{\cal H}\to{\cal
H}$ satisfying
    \be
    H^\dagger=\eta H\eta^{-1}.
    \label{pseudo}
    \end{equation}
Let $\eta:{\cal H}\to{\cal H}$ be such an operator and consider
$\bbr~,~\kkt_\eta:{\cal H}^2\to\C$ defined by
    \be
    \bbr\psi_1|\psi_2\kkt_\eta:=\br\psi_1|\eta\psi_2\kt.
    \label{ps-inner}
    \end{equation}

Clearly, $\bbr~,~\kkt_\eta$ is a nondegenerate Hermitian
sesquilinear form \cite{kato}, i.e., it is a possibly indefinite
inner product --- a pseudo-inner product --- on ${\cal H}$. A
pseudo-Hermitian operator together with a given operator $\eta$
satisfying~(\ref{pseudo}) is said to be $\eta$-pseudo-Hermitian.

The term `pseudo-Hermitian' was introduced in \cite{p1}. But it
turns out that mathematicians \cite{ba} had developed similar
concepts in the study of vector spaces with an indefinite metric,
and Pauli \cite{pauli} had made use of these concepts in his study
of a formulation of the quantum electrodynamics due to Dirac
\cite{dirac}. Note however that there is a seemingly unimportant
but actually quite significant difference between the approach
pursued in the context of spaces with an indefinite metric
(including Pauli's generalization of quantum mechanics to such
spaces) and the point of view adopted  in \cite{p1}. While in the
former one considers a space with a given $\eta$, in the latter
one formulates the concept of pseudo-Hermiticity without having to
specify a particular $\eta$. In fact, as emphasized in \cite{p4}
for a given pseudo-Hermitian Hamiltonian, $\eta$ is not unique.

The basic properties of pseudo-Hermitian operators are the
following \cite{p1,p2,p3,jmp-03,cqg}.
    \begin{itemize}
    \item[] {\bf Theorem~I:} $H$ is $\eta$-pseudo-Hermitian if and only if it is Hermitian with respect to the
pseudo-inner product $\bbr~,~\kkt_\eta$, i.e., for all $\psi_1,\psi_2\in{\cal H}$, $\bbr\psi_1|H\psi_2\kkt_\eta=\bbr H\psi_1|\psi_2\kkt_\eta.$
    \item[] {\bf Theorem~I$\!$I:} Let $H$ be the Hamiltonian of a quantum system and $\eta$ be a linear, Hermitian, invertible operator. Suppose that $\eta$ is time-independent, then $H$ is $\eta$-pseudo-Hermitian if and only if the pseudo-inner product $\bbr~,~\kkt_\eta$ is a dynamical invariant. That is given any two solutions $\psi_1(t)$ and
$\psi_2(t)$ of the Schr\"odinger equation, $i\hbar d\psi/dt= H\psi$,  $\bbr\psi_1(t)|\psi_2(t)\kkt_\eta$ does not depend on time. If $\eta$ depends on time, the pseudo-Hermiticity of $H$ implies
    \be
    \frac{d}{dt}\,\bbr\psi_1(t)|\psi_2(t)\kkt_{\eta(t)}=
    \br\psi_1(t)|\frac{d\eta(t)}{dt}\:\psi_2(t)\kt=
    \bbr\psi_1(t)|\eta^{-1}(t)\:\frac{d\eta(t)}{dt}\:\psi_2(t)\kkt_{\eta(t)}.
    \label{t-dep}
    \end{equation}
\item[] {\bf Theorem~I$\!$I$\!$I:} Suppose $H$ is a diagonalizable Hamiltonian with a discrete spectrum.
Then the following are equivalent.
    \begin{enumerate}
    \item The eigenvalues of $H$ are either real or come in complex-conjugate pairs. In this case we shall say that
    $H$ has a pseudo-real spectrum.
    \item $H$ is pseudo-Hermitian.
    \item $H$ admits an antilinear symmetry generated by an invertible antilinear operator ${\cal X}$, i.e.,    $[H,{\cal X}]=0$.
    \end{enumerate}
\item[] {\bf Theorem~I$\!$V:} Suppose $H$ is a diagonalizable Hamiltonian with a discrete spectrum.
Then the following are equivalent.
\begin{enumerate}
    \item $H$ has a real spectrum,
    \item $H$ is $O^\dagger O$-pseudo-Hermitian for an invertible operator $O$.
    \item $H$ is related to a Hermitian operator by a similarity transformation. Following \cite{quasi}, in this case   $H$ is said to be quasi-Hermitian.
    \item $H$ is Hermitian with respect to a positive-definite inner product.
    \end{enumerate}
\end{itemize}
By definition an inner product $\bbr~,~\kkt$ is said to be
positive-definite if for all nonzero $\psi\in{\cal H}$,
$\bbr\psi|\psi\kkt>0$. In this case $\psi$ is called a positive
vector. Similarly, $\psi$ is called nonnegative, if
$\bbr\psi|\psi\kkt\geq 0$.

As elucidated in \cite{p4}, for a given pseudo-Hermitian
diagonalizable Hamiltonian $H$ the linear, Hermitian, invertible
operators $\eta$ that make $H$ $\eta$-pseudo-Hermitian are, up to
the choice of the eigenbasis of $H$, classified by a set of signs
$\sigma_{n_0} $; $\eta$  has the general form
    \be
    \eta=\sum_{n_0} \sigma_{n_0} |\phi_{n_0}\kt\br\phi_{n_0}|+\sum_{n+}     (|\phi_{n+}+\kt\br\phi_{n-}|
    +|\phi_{n-}\kt\br\phi_{n+}|),
    \label{eta}
    \end{equation}
where $n_0$, $n+$ and $n-$ are spectral labels associated with
eigenvalues with zero, positive, and negative imaginary parts,
$|\phi_n\kt$, with $n=n_0, n+,n-$, are the eigenvectors of
$H^\dagger$ that together with the eigenvectors $|\psi_n\kt$ of
$H$ form a complete biorthonormal system, i.e., they satisfy
    \be
    \br\phi_m|\psi_n\kt=\delta_{mn},~~~~~~~~~~~~ \sum_n |\psi_n\kt\br\phi_n|=1.
    \label{bi}
    \end{equation}
Next, observe that in view of (\ref{eta}) and (\ref{bi}), we have
$\bbr\psi_{n_0}|\psi_{n_0}\kkt_{\eta}=\sigma_{n_0}$, and
$\bbr\psi_{n\pm}|\psi_{n\pm}\kkt_{\eta}=0$. Therefore, the
eigenvectors with complex eigenvalues have zero pseudo-norm; they
are null vectors. Furthermore, the choice $\sigma_{n_0}=+$ for all
$n_0$ implies that all the basis vectors $|\psi_n\kt$ are
nonnegative. In particular, if the spectrum is real this choice
for the signs $\sigma_{n_0}$ yields a basis for the Hilbert space
consisting of positive vectors. This in turn implies that the
inner product $\bbr~|~\kkt_{\eta}$ is positive-definite,
\cite{ba}. This is precisely the inner product whose existence is
ensured by Theorem I$\!$V.

\begin{itemize}
\item[] {\bf Theorem~V:} Suppose $H$ is a pseudo-Hermitian Hamiltonian with a discrete spectrum. Then the most general inner product with respect to which $H$ is Hermitian has the form $\bbr~,~\kkt_{\tilde\eta}$, where $\tilde\eta:=A^\dagger\eta A$, $A$ is an invertible linear operator commuting with $H$, and $\eta$ is given by (\ref{eta}). In particular, if the spectrum of $H$ is real, then the most general positive-definite inner product with respect to which $H$ is Hermitian has the form $\bbr~,~\kkt_{\tilde\eta_+}$, where
    \bea
    \tilde\eta_+&:=&A^\dagger\eta_+ A,
    \label{gen-eta+}\\
    \eta_+&:=&\sum_{n}|\phi_{n}\kt\br\phi_{n}|.
    \label{eta+}
    \eea
Furthermore, if $H$ is a time-independent Hamiltonian with a real
discrete spectrum. Then $\bbr~,~\kkt_{\tilde\eta_+}$ is the most
general invariant positive-definite inner product on the Hilbert
space.
\end{itemize}

Here and also in \cite{p1,p2,p3,p4} we have given the relevant
formulas for the cases that the spectrum of $H$ is discrete. If
the spectrum happens not to be discrete we treat the spectral
label $n$ as a continuous variable, replace the summations with
integrations, and change the Kronecker deltas to Dirac deltas.

\section{Invariant Inner Products for Klein-Gordon Fields}

Consider the Klein-Gordon equation
    \be
    -\ddot\psi(\vec x,t)+\nabla^2\psi(\vec x,t)=\mu^2\psi(\vec x,t),
    \label{kg}
    \end{equation}
where a dot means a derivative with respect to $x^0:=c\,t$, $c$ is
the velocity of light, $\mu:=m\,c/\hbar$, and $m$ is the mass of
the Klein-Gordon field $\psi:\R^{3+1}\to\C$. We can
express~(\ref{kg}) in the form
    \be
    \ddot\psi(\vec x,t)+D\psi(\vec x,t)=0,
    \label{kg2}
    \end{equation}
where $D:=-\nabla^2+\mu^2$. Now, introducing the two-component
state vector $\Psi$ and the (effective) Hamiltonian $H$,
    \be
    \Psi:=\left(\begin{array}{c}
    \psi+i\lambda\dot\psi\\
    \psi-i\lambda\dot\psi\end{array}\right),~~~~~~~~~~
    H:=\frac{1}{2}\left(\begin{array}{cc}
    \lambda D+\lambda^{-1}&\lambda D-\lambda^{-1} \\
    -\lambda D+\lambda^{-1}&-\lambda D-\lambda^{-1}\end{array}\right),
    \label{p-h}
    \end{equation}
with $\lambda$ being an arbitrary nonzero real parameter, we may
express the Klein-Gordon equation in the Schr\"odinger form
$i\dot\Psi= H\Psi$, \cite{fv,G}. It is not difficult to solve the
eigenvalue problem for the Hamiltonian $H$, \cite{jpa-98}. The
eigenvectors $\Psi_{\vec k}$ and the corresponding eigenvalues
$E_{\vec k}$ are given by
    \be
    \Psi_{\vec k}=\left(\begin{array}{c}
    \lambda^{-1}+E_{\vec k}\\
    \lambda^{-1}-E_{\vec k}\end{array}\right)\phi_{\vec k},~~~~~~~~~~
    E_{\vec k}=\pm\sqrt{k^2+\mu^2},
    \label{eg-value}
    \end{equation}
where $\phi_{\vec k}:=\br\vec x|\vec k\kt=(2\pi)^{-3/2}e^{i\vec
k\cdot\vec x}$ and $\vec k\in\R^3$.

The Hamiltonian $H$ of (\ref{p-h}) is not Hermitian with respect
to the $L^2$ inner-product on the space of two-component state
vectors. It is however easy to check that $H$ is
$\sigma_3$-pseudo-Hermitian where $\sigma_3$ is the Pauli matrix
diag$(1,-1)$. In view of Theorem~I$\!$I$\!$I, the
pseudo-Hermiticity of $H$ was to be expected as it is
diagonalizable and has a real spectrum. The pseudo-inner product
$\bbr~,~\kkt_{\sigma_3}$ is nothing but the well-known
Klein-Gordon inner product \cite{fv}, for one can easily check
that $\bbr\Psi_1,\Psi_2{\kkt}_{\sigma_3}=2i\lambda
(\psi_1^*\dot\psi_2-\psi_2^*\dot\psi_1)$. Here the two-component
state vectors $\Psi_i$ are related to one-component state vectors
$\psi_i$ according to (\ref{p-h}). The invariance of the
Klein-Gordon inner product may therefore be viewed as a
manifestation of Theorem~I$\!$I. The much more interesting
observation is that according to Theorem~I$\!$V, $H$ must be
$\eta_+$-pseudo-Hermitian for a positive $\eta_+$ of the form
$O^\dagger O$. The corresponding pseudo-inner product is in fact a
positive-definite inner product. Then according to Theorems~I and
I$\!$I, $H$ is Hermitian with respect to this new
positive-definite inner product and that this inner product is
invariant provided that $\eta_+$ does not depend on time. It is
this inner product that we wish to construct for the effective
Hamiltonian $H$ of (\ref{p-h}).

Having obtained the eigenvectors $\Psi_{\vec k}$ of this
Hamiltonian, we can easily compute the biorthonormal dual vectors
$\Phi_{\vec k}$ and use (\ref{eta+}) to obtain the positive
operator $\eta_+$,
    \bea
    \Phi_{\vec k}&=&\frac{1}{4}\left(\begin{array}{c}
    \lambda +E_{\vec k}^{-1}\\
    \lambda -E_{\vec k}^{-1}\end{array}\right)\phi_{\vec k},\nn\\
    \eta_+&=&\frac{1}{8}\int dk^3
    \left(\begin{array}{cc}
    \lambda^2+ (k^2+\mu^2)^{-1}&\lambda^2- (k^2+\mu^2)^{-1} \\
    \lambda^2-(k^2+\mu^2)^{-1}&\lambda^2+ (k^2+\mu^2)^{-1}
    \end{array}\right)|\vec k\kt\br\vec k|\nn\\
    &=&\frac{1}{8}\left(\begin{array}{cc}
    \lambda^2+ D^{-1}&\lambda^2- D^{-1} \\
    \lambda^2-D^{-1}&\lambda^2+D^{-1}\end{array}\right).
    \label{eta-KG}
    \eea
This in turn implies
    {\small
    \be
    \lambda^{-2}\bbr\Psi_1|\Psi_2\kkt_{\eta_+}=\lambda^{-2}\br\Psi_1|\eta_+\Psi_2\kt=
    \frac{1}{2}
    \left( \int dx^3 \psi_1(\vec x,t)^* \psi_2(\vec x,t)+
    \int dk^3\,\frac{\br\dot\psi_1|\vec k\kt\br\vec k|\dot\psi_2\kt}{k^2+\mu^2}\right)
    =:\cbr\psi_1,\psi_2\ckt,
    \label{1}
    \end{equation}
    }%
where we have denoted the $L^2$-inner product on the space of
two-component state vectors by $\br~|~\kt$, made use of the first
equation in (\ref{p-h}) and (\ref{eta-KG}), and introduced the
inner product $\cbr~,~\ckt$ on the set of solutions of the
Klein-Gordon equation~(\ref{kg}). As seen from (\ref{1}),
$\cbr~,~\ckt$ is a positive-definite inner product. Furthermore,
according to (\ref{eta-KG}) $\eta_+$ is time-independent.
Therefore, in view of Theorem~I$\!$I, $\bbr~|~\kkt_{\eta_+}$ and
consequently $\cbr~,~\ckt$ are dynamical invariants.

One can perform the Fourier integral in (\ref{1}) and obtain
    \be
    \cbr\psi_1,\psi_2\ckt=\frac{1}{2}\,
    \left( \int dx^3 \psi_1(\vec x,t)^* \psi_2(\vec x,t)
    +\int dx^3\int dy^3 \, \dot\psi_1(\vec x,t)^* G(\vec x-\vec y) \dot \psi_2(\vec y,t)\right),
    \label{2}
    \end{equation}
where $G(\vec u):=\exp(-\mu|\vec u|)/(4\pi|\vec u|)$ is a Green's
function for $D$. Indeed, it is not difficult to see that
according to (\ref{1}),
    \be
    \cbr\psi_1,\psi_2\ckt=\frac{1}{2}\,
    \left( \br\psi_1|\psi_2\kt+\br\dot\psi_1|D^{-1}|\dot\psi_2\kt\right),
    \label{3}
    \end{equation}
where $\br~|~\kt$ is the usual $L^2$-inner-product,
$\br\psi_1|\psi_2\kt:=\int dx^3 \psi_1(\vec x,t)^* \psi_2(\vec
x,t)$. The expression (\ref{3}) for the inner-product
$\cbr~,~\ckt$ is quite convenient as it is manifestly
positive-definite and invariant; taking the time-derivative of the
right-hand side of (\ref{3}) and using the Klein-Gordon equation
(\ref{kg}) one finds zero.

Next, we make use of Theorem V to obtain a wide class of
positive-definite inner product $\cbr~,~\ckt$ on the solution
space ${\tilde H}$ of the Klein-Gordon equation. Then, after a
lengthy calculation \cite{cqg}, we find
    \be
    \cbr\psi_1,\psi_2\ckt=\frac{1}{2}\,\left[\br\psi_1|L_+|\psi_2\kt+
    \br\dot\psi_1|L_+D^{-1}|\dot\psi_2\kt+i\left(\br\psi_1|L_-D^{-1/2}|\dot\psi_2\kt-
    \br\dot\psi_1|L_-D^{-1/2}|\psi_2\kt\right)\right],
    \label{gen}
    \end{equation}
where $L_\pm$ are Hermitian linear operators acting in ${\tilde
H}$ such that $A_\pm=L_+\pm L_-$ are positive operators commuting
with $D$.

Expression~(\ref{gen}) has two important properties. First, one
can impose the physical condition of relativistic (Lorentz)
invariance and find out that there is a one-parameter family of
the inner products (\ref{gen}) that are relativistically
invariant. And that the inner product obtained by Woodard in
\cite{woodard} belongs to this family, \cite{cqg}. Second, one can
explore the nonrelativistic limit of these inner products and show
that they indeed tend to the $L^2$-inner product in this limit.
This is a clear indication (besides the invariance and positivity
properties) that one can use the inner-product $\cbr~,~\ckt$ to
devise a probability interpretation for the Klein-Gordon fields.

Furthermore, one can proceed along the lines suggested in
Theorem~I$\!$V and similarity transform the Hamiltonian
(\ref{p-h}) to a Hamiltonian that is Hermitian with respect to the
$L^2$-inner-product on the space of two-component state vectors.
This provides the passage from the pseudo-unitary (or rather
quasi-unitary) quantum mechanics defined by the Hamiltonian
(\ref{p-h}) to the ordinary unitary quantum mechanics in the
(spinorial) Hilbert space $\C^2\otimes L^2(\R^3)$. One can
identify the observables in the latter and use the inverse of the
similarity transformation to define the observables for the
two-component Klein-Gordon fields. These are the
$\eta_+$-pseudo-Hermitian linear operators acting on the space of
two-component state vectors. Because $\eta_+$ is a positive
operator these observables have a real spectrum. Alternatively,
one may define the Hilbert space ${\cal H}_{KG}$ of
single-component Klein-Gordon fields as the set of solutions
$\psi:\R^{3+1}\to\C$ of (\ref{kg}), equivalently the set of
initial conditions, that have finite norm:
$\cbr\psi|\psi\ckt^{1/2}<\infty$, and identify the observables as
linear operators $O:{\cal H}_{KG}\to{\cal H}_{KG}$ that are
Hermitian with respect to the inner product $\cbr~|~\ckt$. It
should be emphasized that in the former approach neither the
passage from the single to two-component state vectors nor the
similarity transform that maps $H$ to a Hermitian Hamiltonian is
unique. As the physical content lies in the original
single-component Klein-Gordon fields $\psi$, one must only
consider the observables whose expectation values are uniquely
determined in terms of $\psi$ (and $\dot\psi$). These will be the
physical observables of the true quantum mechanics of Klein-Gordon
fields and coincide with the linear Hermitian operators acting in
${\cal H}_{KG}$. Note also that the nonuniqueness of the
two-component form of the Klein-Gordon fields that we have
considered manifests in the presence of the arbitrary parameter
$\lambda$. As we have argued in \cite{jpa-98}, this arbitrariness
may be identified with a nonphysical gauge freedom.

An explicit construction of the observables for the Klein-Gordon
fields is provided in \cite{p57}. A comprehensive treatment of the
most general positive-definite inner product on the solution space
of the more general class of Klein-Gordon-type fields, the fact
that all these inner product are unitarily equivalent and define
the same Hilbert space structure, and the nature of the
observables for these fields are discussed in \cite{p54}.

\section{Hilbert Space Problem for Minisuperspace Wheeler-DeWitt Equation}

Consider the Wheeler-DeWitt equation for a FRW minisuperspace
model with a real massive scalar field,
    \be
    \left[ -\frac{\partial^2}{\partial\alpha^2}+\frac{\partial^2}{\partial\varphi^2}+\kappa\,e^{4\alpha}-m^2\,
    e^{6\alpha}\varphi^2\right]\,\psi(\alpha,\varphi)=0,
    \label{wdw}
    \end{equation}
where $\alpha:=\ln a$, $a$ is the scale factor, $\varphi$ is a
real scalar field of mass $m$, $\kappa=-1,0,1$ determines whether
the FRW model describes an open, flat, or closed universe,
respectively, and we have chosen a particularly simple factor
ordering and the natural units, \cite{page,wiltshire}. The
Wheeler-DeWitt equation~(\ref{wdw}) is clearly a Klein-Gordon-type
equation in $1+1$ dimensions. It can be written in the
form~(\ref{kg2}), if we identify a derivative with respect to
$\alpha$ by a dot and let
    \be
    D:=-\frac{\partial^2}{\partial\varphi^2}+m^2\, e^{6\alpha}\varphi^2-\kappa\,e^{4\alpha}.
    \label{D=}
    \end{equation}
In view of this identification we shall take $\alpha$ as the
time-coordinate. Moreover, we can use (\ref{p-h}) to obtain a
two-component formulation for the Wheeler-DeWitt
equation~(\ref{wdw}). Again the corresponding effective
Hamiltonian is diagonalizable. Its eigenvectors $\Psi_{n\pm}$ and
the corresponding eigenvalues $E_{n\pm}$ are given by
\cite{jmp-98}
    \be
    \Psi_{n\pm}=\left(\begin{array}{c}
    \lambda^{-1}+E_{n\pm}\\
    \lambda^{-1}-E_{n\pm}\end{array}\right)\phi_{n},~~~~~~~~~~
    E_{n\pm}=\pm\sqrt{m\,e^{3\alpha}(2n+1)-\kappa\,e^{4\alpha}},
    \label{eg-value2}
    \end{equation}
where $n=0,1,2,\cdots$, $\phi_{n}:=\br\varphi|n\kt=N_n
H_n(m^{1/2}e^{3\alpha/2}\varphi)\, e^{-m\,e^{3\alpha}\varphi^2/2}$
are the energy eigenfunctions of a simple harmonic operator with
unit mass and frequency $m\,e^{3\alpha}$, $H_n$ are Hermite
polynomials, and $N_n:=[m\,e^{3\alpha}/(\pi 2^{2n}{n!}^2)]^{1/4}$
are normalization constants.

As seen from (\ref{eg-value2}), the spectrum of $H$ is discrete
and pseudo-real. Hence according to Theorem~I$\!$I$\!$I, it is
pseudo-Hermitian, \cite{p1}. In particular, it is
$\sigma_3$-pseudo-Hermitian. The indefinite inner-product
$\bbr~,~\kkt_{\sigma_3}$ is the invariant Klein-Gordon inner
product that is often used in the probability interpretation of
the semiclassical Wheeler-DeWitt fields,
\cite{bryce-1,vilenkin,wald,parentani}. Also note that for the
open and flat universes the spectrum of $H$ is real. Therefore,
$H$ is quasi-Hermitian and there is a positive linear Hermitian
invertible operator $\eta_+$ such that $H$ is
$\eta_+$-pseudo-Hermitian. In fact, $\eta_+$ is given by
Eq.~(\ref{eta-KG}) where $D$ has the form~(\ref{D=}). Following
the above treatment of the Klein-Gordon equation, we can use this
$\eta_+$ to obtain a positive-definite inner product on the space
of the Wheeler-DeWitt fields. The latter is given by Eq.~(\ref{3})
where the $L^2$-inner product has the form
$\br\psi_1|\psi_2\kt:=\int_{-\infty}^\infty d\varphi
\psi_1(\alpha,\varphi)^*\psi_2(\alpha,\varphi)$.  Eq.~(\ref{3})
involves the inverse of $D$ which is well-defined for the open and
flat universes where $\kappa=-1$ or $0$. For these cases,
$D^{-1}=\sum_{n=0}^\infty
[m\,e^{3\alpha}(2n+1)-\kappa\,e^{4\alpha}]^{-1}|n\kt\br n|$. For
the closed universe the above construction works only for
$a=e^{\alpha}< m$ where all the eigenvalues are real. For $a>m$,
the complex-conjugate imaginary eigenvalues are also present and
the corresponding eigenvectors will be null.

For the cases that the spectrum is real ($H$ is quasi-Hermitian),
we can follow the statement of Theorem~I$\!$V to perform a
similarity transformation to map the effective Hamiltonian to a
Hamiltonian that is Hermitian in the $L^2$-inner product on the
space of two-component state vectors, i.e., $\C^2\otimes L^2(\R)$.
This is in complete analogy with the case of Klein-Gordon
equation~(\ref{kg}). However, there is an important distinction
between the Klein-Gordon equation~(\ref{kg}) and the
Wheeler-DeWitt equation~(\ref{wdw}), namely that for the latter
equation the operators $D$ and $\eta_+$ are `time-dependent.' This
in particular means that the associated positive inner product
$\cbr~,~\ckt$ is not invariant. Instead, as a consequence of
(\ref{t-dep}) or alternatively (\ref{3}) and (\ref{kg2}), it
satisfies,
    \[\frac{d}{d\alpha}\cbr\psi_1,\psi_2\ckt=\frac{1}{2}\, \br\psi_1|
    \frac{d(D^{-1})}{d\alpha}|\psi_2\kt.\]
Note that the similarity transformation that maps $H$ into a
Hermitian effective Hamiltonian is `time-dependent' as well. This
makes it fail to preserve the form of the associated Schr\"odinger
equation, i.e., it is not a pseudo-canonical transformation
\cite{p3}. Therefore, although this similarity transformation may
be used to related the observables of the two systems, they do not
relate the solutions  of the corresponding Schr\"odinger
equations.

In Ref.~\cite{cqg}, we have outlined a method to define a unitary
quantum evolution for the cases that the Hilbert space has a
time-dependent inner product structure. Using this method and the
above results for time-independent Hamiltonians, one can obtain
the general form of an invariant positive-definite inner product
on the solution space of the Wheeler-DeWitt equation (\ref{wdw})
for the case of a flat or open universe or when the initial value
of the scale factor $a$ is less than $m$. A rather comprehensive
treatment is provided in \cite{p54}.

\section*{Acknowledgment}
This work has been supported by the Turkish Academy of Sciences in
the framework of the Young Researcher Award Program
(EA-T$\ddot{\rm U}$BA-GEB$\dot{\rm I}$P/2001-1-1).

\ed